\documentclass[12pt]{article}
\usepackage{amsmath}
\usepackage{amssymb}
\begin{document}
\vskip 2cm
\begin{center}
{\bf {\Large Superspace Unitary Operator in Superfield Approach to 
Non-Abelian Gauge Theory with Dirac Fields}}\\

\vskip 3.2cm

{\sf T. Bhanja$^{(a)}$, D. Shukla$^{(a)}$, R. P. Malik$^{(a,b)}$}\\
$^{(a)}$ {\it Physics Department, Institute of Science,}\\
{\it Banaras Hindu University, Varanasi - 221 005, (U.P.), India}\\

\vskip 0.1cm


\vskip 0.1cm

$^{(b)}$ {\it DST Centre for Interdisciplinary Mathematical Sciences,}\\
{\it Institute of Science Banaras Hindu University, Varanasi - 221 005, (U.P.), India}\\
{\small {\sf {E-mails:  tapobroto.bhanja@gmail.com; dheerajkumarshukla@gmail.com;
rpmalik1995@gmail.com}}}

\end{center}

\vskip 3cm

\noindent
{\bf Abstract:} Within the framework of  augmented version of the superfield
approach to Becchi-Rouet-Stora-Tyutin (BRST) formalism, we derive the superspace 
unitary operator (and its Hermitian conjugate) in the context of four (3+1)-dimensional 
(4$D$) interacting non-Abelian 1-form gauge theory with Dirac fields. The ordinary 4$D$ non-Abelian 
theory, defined on the flat 4$D$ Minkowski spacetime manifold, is generalized onto a 
(4, 2)-dimensional supermanifold which is parameterized by the spacetime bosonic 
coordinates $x^\mu$ (with $\mu = 0, 1, 2, 3$) and a pair of Grassmannian variables 
($\theta, \bar\theta$) which satisfy the standard relationships: $\theta^2 = 
{\bar\theta}^2 = 0, \theta\,\bar\theta + \bar\theta\,\theta = 0$.
Various consequences of the application of the above superspace (SUSP) unitary operator 
(and its Hermitian conjugate) are discussed. In particular, we obtain the results of 
the application of horizontality condition (HC) and gauge invariant restriction 
(GIR) in the language of the above SUSP operators. One of the novel results of our present
 investigation is the derivation of explicit expressions for
the SUSP unitary operator (and its Hermitian conjugate) without imposing any Hermitian conjugation
condition from {\it outside} on the parameters and (super)fields of the supersymmetric version of our
4$D$ interacting non-Abelian 1-form theory with Dirac fields. \\

\vskip 1.2cm



\newpage
\noindent
\section{Introduction}

The covariant canonical quantization of a given gauge theory is performed within
the framework of Becchi-Rouet-Stora-Tyutin (BRST) formalism where the local gauge
symmetries of the {\it classical} gauge theory are traded with the quantum gauge (i.e. BRST)
symmetries. For a single classical local gauge symmetry, there exist two {\it quantum} gauge 
symmetries which are christened as the BRST and anti-BRST symmetries. In one stroke,
the BRST formalism provides (i) the covariant canonical quantization, (ii) the proof of  unitarity, 
(iii) the physicality criteria in the quantum Hilbert space, etc., for a given gauge theory. 
The two key mathematical properties, associated with the above (anti-)BRST symmetries, 
are the nilpotency property and absolute anticommutativity. The former property 
establishes the fermionic   (i.e. supersymmetric) nature of these symmetries 
and the latter property encodes the linear independence of the BRST and anti-BRST
transformations.

The superfield approach (see, e.g. [1-5]) to BRST formalism provides the 
geometrical origin and interpretation for the nilpotency and absolute anticommutativity 
properties of the (anti-)BRST symmetries that are required for the covariant canonical 
quantization of the  $p$-form ($p = 1, 2, 3,...$) gauge theories. In particular, the 
horizontality condition (HC) plays a pivotal role in the derivation of the nilpotent (anti-)BRST 
symmetry transformations connected with the $p$-form Abelian gauge and corresponding (anti-)ghost 
fields of the theory and it also leads to the systematic derivation of the 
Curci-Ferrari (CF) condition(s). The latter have been shown to be connected with 
the geometrical objects called gerbes in the context of  BRST description of the Abelian 
2-form and Abelian 3-form gauge theories [6-7] in {\it four} (3 + 1)- and {\it six} (5 + 1)-dimensions 
of the flat Minkowskian spacetime, respectively.

The above usual superfield approach [1-5] has been systematically generalized to
incorporate the {\it additional} restrictions on the superfields so as to derive  
the proper (anti-)BRST symmetries for the matter fields in an {\it interacting} gauge theory (in 
addition to the nilpotent (anti-)BRST symmetries for the gauge and (anti-)ghost fields that 
are derived due to HC). These additional restrictions are found to be consistent with the HC 
because the geometrical interpretations for the (anti-)BRST symmetries remain intact. 
In particular, it is the gauge-invariant restrictions involving the covariant derivatives on 
the matter fields which play a decisive role in this direction where the inputs from the HC 
also plays a key role. We have christened these generalized versions of superfield approach to BRST 
formalism as the augmented version of superfield formalism [8-10]. The latter
(i.e. the augmented version of superfield formalism) are basically the consistent 
generalizations of the usual superfield formalism (particularly developed in [1-3]).

A superspace (SUSP) unitary operator has been shown to exist in the superfield 
approach to the description of 4$D$ non-Abelian 1-form gauge theory within the framework
of BRST formalism (see, e.g. [1-3]). The beauty of this operator is the observation 
that it maintains the explicit $SU(N)$ group structure in the transformation space 
of the superspace (where the ordinary fields, defined on the $D$-dimensional ordinary 
flat Minkowskian manifold, are generalized onto the superfields defined on the the 
($D$, 2)-dimensional supermanifold). This explicit group structure is somewhat hidden in the 
{\it direct} application of HC. Thus, the utility of the SUSY unitary operator has an edge 
over the utility of HC because the results of the latter could be derived by 
using the former (cf. Sec. 5). The central theme of our present paper is to derive 
this unitary operator (and its Hermitian conjugate) explicitly by applying the augmented 
version of superfield formalism [8-10] in the description of the 4$D$ non-Abelian
interacting theory with Dirac fields. It should be noted that this operator has been
judiciously chosen in [1-3] for the BRST description of the 4$D$ interacting non-Abelian
gauge theory with a {\it generic} matter field. One of the central aims of our present 
investigation is to theoretically prove the sanctity and preciseness  of this choice.

Our present investigation is motivated by the following key factors. First, 
the explicit computation of the SUSP unitary operator $U (x, \theta, \bar\theta)$
in $\psi (x) \rightarrow \Psi(x, \theta, \bar\theta) = U (x, \theta, \bar\theta)
\,\psi (x)$ maintains the $SU(N)$ group structure in the transformation space.
Second, the existence of  this $SU(N)$ group structure leads to the definition of 
the super covariant derivative on the superfield which obeys the same transformation
property(i.e. $D\psi \to \, {\tilde D}\Psi = U (x, \theta, \bar\theta)\,
D\psi $). Third, the definition of the covariant derivative and its transformation rule
yield the transformation rule for the super curvature 2-form (i.e. $D\,D\,\psi 
\to\,{\tilde D}\,{\tilde D}\,\Psi = U (D\,D\,\psi) \,\Longrightarrow\,
{\tilde F}^{(2)} = U\, F^{(2)}\,U^\dagger$). Fourth, the above arguments lead to the 
results that are obtained {\it only} due to the application of HC (cf. Sec. 3 below). Thus, in some sense, 
the derivation of the SUSP operators $U (x, \theta, \bar\theta)$ and $U^\dagger 
(x, \theta, \bar\theta)$ provides an alternative to the HC where $SU(N)$ group 
structure is respected very {\it clearly}. Fifth, we theoretically derive the SUSP
unitary operator and mathematically support the sanctity of the choice made in [1-3].
The judicious choice made in Ref. [1-3] is somewhat straightforward when one knows the (anti-)BRST 
symmetry transformations before hand. However, the explicit derivation, within the framework 
of our superfield formalism, is a novel result for the $4D$ non-Abelian 1-form gauge theory 
with Dirac fields.  Finally, our present attempt is the generalization of our earlier 
work on the 4$D$ {\it Abelian} 1-form interacting gauge theory with Dirac and complex scalar 
fields [11] to the 4$D$ interacting {\it non-Abelian} 1-form gauge theory with Dirac fields.

The contents of our present investigation are organized as follows. To set-up the
notations and convention, in Sec. 2, we briefly recapitulate the bare essentials
of the (anti-)BRST symmetries for the 4$D$ interacting  non-Abelian gauge theory 
with Dirac fields in the Lagrangian formulation. Our Sec. 3 is devoted to a concise 
discussion on the horizontality condition (HC) and the derivation of the nilpotent (anti-)BRST 
symmetries for the gauge and (anti-) ghost fields. In Sec. 4, we derive the nilpotent (anti-)BRST 
symmetry transformations for the Dirac fields by exploiting the power and potential 
of gauge invariant restriction (GIR) on the superfields defined on the (4, 2)-dimensional 
supermanifold. Our Sec. 5 contains the derivation of SUSP unitary operator and it 
incorporates  various consequences that emerge out from the knowledge of 
this unitary operator (and its Hermitian conjugate). Finally, in our Sec. 6, we summarize 
our key results and point out a few future directions for further investigations.
In our Appendix A, we discuss a few things for readers' convenience.\\

\noindent
{\it Convention and Notations:} We adopt the convention and notations such
that the 4$D$ Minkowskian flat spacetime manifold is endowed with a metric $\eta_{\mu\nu}$
with signatures $(+ 1, - 1, - 1, - 1)$ so that 
$\partial_\mu \, A^\mu = \eta_{\mu\nu} \partial^\mu \, A^\nu = \partial_0 \, A_0 -
\partial_i \, A_i $ where the Greek indices $\mu, \, \nu, \, \lambda,... = 0, 1, 2, 3$ correspond
to the spacetime directions and Latin indices $i,\, j,\, k,... = 1, 2, 3$ stand for the space directions
{\it only}. In the Lie-algebraic space, we take the dot and cross products between $P^a$ and $Q^a$ as:
$P \cdot Q = P^a \, Q^a $ and $(P \times Q)^a = f^{abc} P^b \, Q^c$ where $f^{abc}$  are the 
structure constants that can be chosen to be totally antisymmetric for the $SU(N)$ group
with the Lie algebra $[T^a, \, T^b] = f^{abc}\, T^c$. Here $T^a$ (with $a = 1, 2, 3, ... N^2 - 1) $ 
are the generators of the Lie algebra corresponding to the $SU(N)$ group. 
We have used the same notation for the covariant derivative and the dimensionality 
of the spacetime. However, the different meanings, attached with the symbol $D$, are quite unambiguous in 
the whole body of the text of our present endeavor.
We have explicitly taken the covariant derivatives on the matter field and ghost fields as:
$D_\mu \,\psi = (\partial_\mu + i \,A_\mu \cdot T)\,\psi$ and 
$D_\mu\, C = \partial_\mu \,C + i\, (A_\mu \times C)$ which are in the {\it fundamental}
and {\it adjoint} representations, respectively.

\noindent
\section{Preliminaries: (Anti-)BRST Symmetries}

We begin with the (anti-)BRST invariant Lagrangian densities ${\cal L}_B$ and ${\cal L}_{\bar B}$ 
(see, e.g. [12, 13]) for 
the 4$D$ interacting non-Abelian 1-form gauge theory with the {\it massive} (with mass $m$)
Dirac fields ($\psi,\, \bar\psi$) , in the Curci-Ferrari gauge (where $\xi = 2$) [14,15], as
\begin{eqnarray}
{\cal L}_B &=& - \frac{1}{4}\, F_{\mu\nu} \cdot F^{\mu\nu} + \bar \psi\, (i\,\gamma^\mu\,D_\mu - m)\, 
\psi + s_b\,s_{ab}\left({i \over 2}\, A_\mu \cdot A^\mu -\,{\xi  \over 2}\; \bar C\cdot C \right) \nonumber\\
 &\equiv & - \frac{1}{4}\, F_{\mu\nu} \cdot F^{\mu\nu} + \bar \psi\, (i\,\gamma^\mu\,D_\mu - m)\, 
\psi + B\cdot (\partial_\mu A^\mu) + \frac{1}{2} (B\cdot B + \bar B \cdot \bar B)\nonumber\\ 
&-&\,i\,\partial_\mu\bar C \cdot D^\mu C, \nonumber\\
{\cal L}_{\bar B} &=& - \frac{1}{4}\, F_{\mu\nu} \cdot F^{\mu\nu} + \bar \psi\, (i\,\gamma^\mu\,D_\mu - m)\, 
\psi - s_{ab}\,s_b\left({i \over 2}\, A_\mu \cdot A^\mu -\,{\xi \over 2} \, \bar C\cdot C\right) \nonumber\\
 & \equiv & - \frac{1}{4}\, F_{\mu\nu} \cdot F^{\mu\nu} + \bar \psi\, (i\,\gamma^\mu\,
D_\mu - m)\, \psi \, - \bar B \cdot (\partial_\mu A^\mu) + \frac{1}{2} (B\cdot B + \bar B \cdot \bar B) 
\nonumber\\  &-& i D_\mu\bar C \cdot \partial^\mu C,
\end{eqnarray} 
 for the explicit derivation of the gauge-fixing and Faddeev-Popov ghost terms. In other words, we have derived 
the final expressions for the Lagrangian densities ${\cal L}_B$ and ${\cal L}_{\bar B} $ by taking
 into account the Curci-Ferrari gauge where $\xi = 2$. It will be noted 
that  the above Lagrangian densities ${\cal L}_B$ and ${\cal L}_{\bar B} $ are equivalent on 
the constrained hypersurface defined by the field equation: $ B + \bar B = -(C \times \bar C)$ 
which is nothing but the Curci-Ferrari (CF) condition [16]. In the above, we have the 
curvature tensor $F_{\mu\nu} = \partial_\mu A_\nu - \partial_\nu A_\mu 
+ i\, (A_\mu \times  A_\nu)$ which is derived from the 2-from 
$F^{(2)} = d A^{(1)} + i\, A^{(1)} \wedge\, A^{(1)}$ where the connection
1-from $A^{(1)} = dx^\mu (A_\mu \cdot T)$ defines the vector potential $A^{a}_\mu $
in the $SU(N)$ Lie-algebraic space (with $a,b,c,... = 1,2,3...N^2 - 1$).  The fields $B(x)$ 
and $\bar B(x)$ are the Nakanishi-Lautrup auxiliary fields and the fermionic
[$({C^a})^2 = ({\bar C^a})^2 = 0,\; {C^a} \, {\bar C^b} + {\bar C^b}\,{C^a} = 0$, etc.]
(anti-)ghost fields $({\bar C}^a)\, {C^a}$ are required  for the validity of 
unitarity in the theory.

We have a covariant derivative on the Dirac field $\psi$ as:
$D_\mu \, \psi = (\partial_\mu + i \, g\, A_\mu \cdot T) \,\psi \equiv (\partial_\mu 
+ i \, A_\mu \cdot T) \, \psi$ where the coupling constant $g$ has been set equal to 
{\it one} for the sake of brevity. Similarly, we have taken into consideration the definition 
of the covariant derivatives (with  $g = 1$) 
on the (anti-)ghost fields as: $D_\mu \bar C = \partial_\mu \bar C + i\,(A_\mu \times \bar C)$ 
and $D_\mu C = \partial_\mu C + i\,(A_\mu \times C)$ .  The Dirac fields are fermionic ($\psi^2 
= {\bar\psi}^2 = 0,\, \psi\,\bar\psi + \bar\psi\,\psi = 0$) in nature because they 
commute ($\psi\, A_\mu - A_\mu \, \psi = 0, \, \psi\, B - B\, \psi = 0,$ etc.) with all 
the bosonic fields (e.g. $A_\mu, \, B, \, \bar B, \, F_{\mu\nu}$) but they anticommute 
($\psi\,  C^a + C^a\, \psi = 0, \, C^a\,\bar\psi + \bar\psi \, C^a = 0,\,{\bar C}^a\, \psi + \psi \, 
{\bar C}^a = 0,$ etc.) with the fermionic (anti-)ghost fields $({\bar C}^a)C^a$ of our theory. The above 
Lagrangian densities (1) respect the following off-shell nilpotent ($s_{(a)b}^2 = 0$), 
absolutely anticommuting ($s_b\,s_{ab} + s_{ab}\,s_b = 0$), continuous and infinitesimal
(anti-)BRST symmetry transformations $s_{(a)b}$ (see, e.g. [17] and our Appendix A for details)
\begin{eqnarray}
&& s_{ab}\,A_\mu = D_\mu\,\bar C, \qquad s_{ab}\,\bar C =  -\,\frac{i}{2}\,(\bar C \times \bar C), 
\qquad s_{ab}\,C = i\,\bar B, \qquad s_{ab}\,F_{\mu\nu} = i\, (F_{\mu\nu} \times \bar C), \nonumber\\ 
&& s_{ab}\,B = i\, (B \times \bar C),\qquad  s_{ab}\,\bar B = 0, 
\qquad\quad s_{ab}\,\psi = -\,i\,(\bar C \cdot T)\,\psi,
\qquad s_{ab}\,\bar\psi = -\,i\,\bar\psi\, (\bar C \cdot T),\nonumber\\
&& s_b\,A_\mu = D_\mu\,C, \qquad s_b\, C =  -\,\frac{i}{2}\,(C \times C), \qquad\quad s_b\,\bar{C} = i\,B,
\qquad  s_b\,F_{\mu\nu} = i\, (F_{\mu\nu} \times C),\nonumber\\ 
&& s_b\,B = 0,  \qquad s_b\,\bar{B} = i\, (\bar{B} \times C),  \qquad\quad s_b\,\psi = -\,i\,(C \cdot T)\,\psi,
 \qquad s_b\,\bar\psi = -\,i\,\bar\psi\, (C \cdot T).
\end{eqnarray}
The noteworthy points, at this stage, are as follows:
(i) The kinetic term $\left[(-1/4)\, F^{\mu\nu}\cdot F_{\mu\nu}\right]$ remains invariant 
under the (anti-)BRST symmetry transformations. Geometrically, the curvature tensor 
$F_{\mu\nu}$ has its origin in the exterior derivative $d$ because $F^{(2)} =  d\, A^{(1)} 
+ i\, A^{(1)} \wedge A^{(1)}$ defines the $F_{\mu\nu }$ tensor (which can be 
explicitly written in terms of the potential $A_\mu$ as:
$F_{\mu\nu} = \partial_\mu A_\nu - \partial_\nu A_\mu + i\,[A_\mu,\, A_\nu]$
where we observe that $F_{\mu\nu} = F_{\mu\nu} \cdot T$ and $A_\mu =  A_\mu \cdot T$). 
(ii) The CF-condition remains invariant (i.e. $s_{(a)b}\,\left[B + \bar B 
+ (C \times \bar C) \right]= 0$)
under the off-shell nilpotent (anti-)BRST symmetry transformations. Thus, 
it is a {\it physical} constraint that can be imposed on the 4$D$ non-Abelian interacting 
1-form gauge theory. (iii) The Lagrangian densities ${\cal L}_B$ and ${\cal L}_{\bar B}$ 
{\it both} are equivalent on the hypersurface in the 4$D$ Minkowskian spacetime manifold 
due to the CF condition (i.e.  $B + \bar B + (C \times \bar C) = 0$).
In fact, it is on this hypersurface that the off-shell nilpotency and
absolute anticommutativity properties of the (anti-)BRST symmetry transformations are
valid (see, e.g. [12-15]) in an explicit manner.

\noindent
\section{Symmetries for the Gauge and (Anti-)Ghost Fields: Horizontality Condition}

In this section, we shall exploit the geometrical beauty of $F^{(2)} = d A^{(1)} 
+ i\, A^{(1)} \wedge A^{(1)} $ in the context of HC within the framework of superfield formalism 
and derive the CF-condition: $B + \bar B + (C \times \bar C) = 0$ as well as
the proper (anti-)BRST symmetry  transformations for the gauge and (anti-)ghost 
fields of the 4D non-Abelian 1-form gauge theory. For this purpose, first of all, 
we define the supercurvature 2-form ${\tilde F}^{(2)}$ on the (4, 2)-dimensional supermanifold 
${\tilde F}^{(2)} =\, \tilde{d}\,\tilde A^{(1)} 
+ i\,\tilde A^{(1)} \wedge \tilde A^{(1)} = \,\frac{1}{2}\,\left(dZ^M \wedge dZ^N\right)\,
{\tilde F}_{MN}(x, \theta, \bar\theta) $ where $Z^M = (x^\mu, \theta, \bar\theta)$ is 
the superspace coordinate that characterizes the (4, 2)-dimensional supermanifold with 
$x^\mu$ ($\mu = 0, 1, 2, 3$) as the bosonic variables and $(\theta, \bar\theta)$ as 
a pair of fermionic $(\theta^2 = {\bar\theta}^2 = 0,\, \theta\,\bar\theta + \bar\theta\,
\theta = 0)$ Grassmannian variables. The other symbols of relevance, on the 
appropriately chosen  (4, 2)-dimensional supermanifold, are  as follows [1-3, 17]
\begin{eqnarray}
&&d = dx^\mu\, \partial_\mu \rightarrow   \tilde d = dZ^M\, \partial_M 
= dx^\mu\, \partial_\mu + d\theta\, \partial_\theta + d\bar\theta\,\partial_{\bar\theta},
\nonumber\\
&&A^{(1)} = dx^\mu\,A_\mu \rightarrow  \tilde A^{(1)} = dZ^M\,A_M 
= dx^\mu\,B_\mu + d\theta\, \bar F + d\bar\theta\, F,
\end{eqnarray}
where $A_M = \left(B_\mu (x, \theta, \bar\theta),\, F (x, \theta, \bar\theta),\,\bar F
(x, \theta, \bar\theta) \right)$ constitutes the vector multiplet and the superspace derivative 
$\partial_M$ stands for $(\partial_\mu, \partial_\theta,\partial_{\bar\theta})$. The latter 
(i.e. $\partial_M$) are the components of the superspace derivative 
($\partial_M = \partial/\partial Z^M $) with $ Z^M = (x^\mu, \theta, \bar\theta)$.

The requirement of the HC is to set all the Grassmannian components of 
${\tilde F}_{MN} (x, \theta, \bar\theta) = \bigl( {\tilde F}_{\mu\theta},\,
{\tilde F}_{\mu\bar\theta},\,{\tilde F}_{\theta \bar\theta},\, {\tilde F}_{\theta \theta},
\,{\tilde F}_{\bar\theta \bar\theta}\bigr)$ equal to zero (i.e. $ {\tilde F}_{\mu\theta} 
= {\tilde F}_{\mu\bar\theta} = {\tilde F}_{\theta \bar\theta}
= {\tilde F}_{\theta \theta} = {\tilde F}_{\bar\theta \bar\theta} = 0$). 
We know that the kinetic term [$-(1/4)\,F^{\mu\nu} \cdot F_{\mu\nu}$] for the 
gauge field is an (anti-)BRST invariant quantity. The HC basically demands the independence of this 
quantity (i.e. $-\, {1\over 4}\,\tilde F_{MN}\cdot \tilde F^{MN} = -\, {1\over 4}\, F_{\mu\nu}\cdot F^{\mu\nu}$). 
In another words, this gauge (i.e. BRST) invariant quantity should be independent of the Grassmannian variables
because the latter cannot be realized physically in the ordinary space.
This condition leads to the derivation of the relationships between the basic as well as auxiliary fields of the 
Lagrangian densities ${\cal L}_B$ and ${\cal L}_{\bar B}$ {\it and} the secondary fields 
$(R_\mu, {\bar R}_\mu, S_\mu, \bar B_1, B_1, \bar B_2, B_2, s, \bar s)$ of the following 
expansions (see, e.g. [1-3] for details)
\begin{eqnarray}
B_\mu  (x, \theta, \bar\theta) &=& A_\mu (x) + \theta\, \bar R_\mu  (x)
+ \bar\theta \, R_\mu (x) + i\,\theta\bar\theta \, S_\mu (x), \nonumber\\
F (x, \theta, \bar\theta) &=& C (x) + i\,\theta\, \bar B_1 (x) + i\,\bar\theta\, B_1 (x)
+ i\,\theta \bar\theta\,s(x), \nonumber\\
\bar F  (x, \theta, \bar\theta) &=& \bar C (x) + i\,\theta\, \bar B_2 (x) 
+ i\, \bar\theta \,B_2 (x) + i\,\theta  \bar\theta\, \bar s (x), 
\end{eqnarray}
where, as pointed out earlier, the superfields $B_\mu (x, \theta, \bar\theta),$
$F (x, \theta, \bar\theta), \bar F  (x, \theta, \bar\theta)$ constitute the  
multiplet of the vector superfield $A_M  (x, \theta, \bar\theta)$.
The above expansions are nothing but the shift transformations for the superfields along the
Grassmannian directions $ (1, \theta, \bar\theta, \theta\bar\theta)$ of the (4, 2)-dimensional 
supermanifold. The HC (i.e. $ {\tilde F}_{\theta \bar\theta} = {\tilde F}_{\theta \theta} 
= {\tilde F}_{\bar\theta \bar\theta} = {\tilde F}_{\mu\theta} = {\tilde F}_{\mu\bar\theta}
= 0$) yields the following interesting and very useful relationships 
(with $\bar B_1 = \bar B, \, B_2 =  B  $) [1-3, 17]
\begin{eqnarray}
&& R_\mu = D_\mu\,C, \quad\qquad {\bar R}_\mu = D_\mu \, \bar C, 
\quad\qquad  B_1 = -\, \frac{1}{2}\, (C \times C),\nonumber\\
&& S_\mu = D_\mu B_2 + D_\mu C \times \bar C \equiv -\,D_\mu\, \bar B_1 - D_\mu\,\bar C \times C, \nonumber\\
&& \bar B_2 = -\,\frac{1}{2}\, (\bar C \times\bar C), \qquad \,\, s = i\, (\bar B_1 \times C), 
\,\,\qquad \bar s = -\,i\,(B_2 \times \bar C), \nonumber\\
&& \bar B_1 + B_2 =  -\,(C \times \bar C)  \rightarrow  B + \bar B = -\, (C \times \bar C), 
\end{eqnarray}
where $D_\mu \,C = \partial_\mu \, C + i\, (A_\mu \times C)$  and the last entry, in the above, is nothing 
but the CF-condition. The substitution of the above expressions for the secondary fields into (4) yields
the following
\begin{eqnarray}
B_\mu^{(h)} (x, \theta, \bar\theta) &=& A_\mu + \theta\,(D_\mu\, \bar C) + \bar\theta\,(D_\mu \,C) 
+ \theta\bar\theta\, \left[i\,\bigl(D_\mu B + \left(D_\mu C \times \bar C\right)\bigr)\right]\nonumber\\
&\equiv &  A_\mu + \theta\, (s_{ab}\, A_\mu) + \bar \theta\, (s_b\, A_\mu) 
+ i\, \theta\bar\theta \, (s_b\,s_{ab}\, A_\mu), \nonumber\\
F^{(h)} (x, \theta, \bar\theta) &=& C (x) + \theta\, (i\, \bar B) + \bar\theta \,(-\,\frac{i}{2}\,C \times C)
+ \theta\bar\theta\, (-\,\bar B \times C) \nonumber\\
&\equiv &  C + \theta\, (s_{ab}\,C) + \bar\theta \, (s_b\,C) + \theta \bar{\theta} 
\,(s_b\,s_{ab}\,C), \nonumber\\
\bar F^{(h)} (x, \theta, \bar\theta) &=&  \bar C (x) + \theta\,(-\,\frac{i}{2}\,\bar{C} \times \bar C)
+ \bar\theta\,(i\,B) + \theta\bar\theta\,(B \times \bar C) \nonumber\\
& \equiv & \bar C (x) + \theta\, (s_{ab}\,\bar C) + \bar\theta \, (s_b\,\bar C) 
+ \theta\bar{\theta} \,(s_b\,s_{ab}\,\bar C),
\end{eqnarray}
where the superscript $(h)$ denotes the expansions obtained after the application of HC.
It is to be noted that the CF-condition $B + \bar B + (C \times \bar C) = 0$ emerges out from 
setting the coefficient of $(d\theta \wedge d \bar\theta)$ equal to {\it zero} in the  
relationship $\tilde d\,\tilde A^{(1)} + i\,\tilde A^{(1)} \wedge \tilde A^{(1)}$ 
which results in the condition $\partial_\theta\, F^{(h)} + \partial_{\bar\theta}\,\bar F^{(h)} 
- i\, \lbrace F^{(h)},\,\bar F^{(h)}\rbrace = 0$. It is elementary to check that the explicit 
substitutions from (6) into {\it this} restriction yields the CF-condition $\left( B + \bar B 
+ (C \times \bar C) = 0 \right)$. The (anti-)BRST invariance of the CF-condition 
can be captured within the framework of superfield formalism as it can be readily checked that: 
$\partial_\theta \big[ \partial_\theta\, F^{(h)} + \partial_{\bar\theta}\,\bar F^{(h)}
- i\, \lbrace F^{(h)},\,\bar F^{(h)}\rbrace \big]|_{\bar\theta = 0 }\; =\; 0$  and 
$\partial_{\bar\theta} \,\bigl[\partial_\theta F^{(h)} + \partial_{\bar\theta} {\bar F}^{(h)}
 -i\, \lbrace F^{(h)}, \,{\bar F}^{(h)}\rbrace\bigr]|_{\theta = 0 }\; =\; 0$ which 
 {\it physically} imply the (anti-)BRST invariance $\bigl(s_{(a)b}\,[B + \bar B + (C \times \bar C)] 
= 0\bigr)$ of the CF-condition defined on the ordinary 4D Minkowskian spacetime manifold as the
constrained field equation.

\noindent 
\section{(Anti-)BRST Symmetries for the Matter Fields: Gauge Invariant Restriction}

We exploit the strength of the gauge invariant restriction (GIR) on the superfields, defined 
on the (4, 2)-dimensional supermanifold, to obtain the proper (anti-)BRST symmetry 
transformations for the matter fields of the {\it interacting} 4$D$ non-Abelian gauge theory 
with Dirac fields. This appropriate condition (which incorporates the results of HC
(cf. (6))), is as follows [17]
\begin{equation}
\bar\Psi (x, \theta, \bar\theta)\, \bigl(\tilde d + i\, {\tilde A}^{(1)}_{(h)}\bigr)\, 
\Psi (x, \theta, \bar\theta)  = \bar\psi (x) \, (d + i\,A^{(1)})\, \psi (x),
\end{equation}
where the super 1-form connection on the l.h.s., derived after the application of HC,
 is: ${\tilde A}^{(1)}_{(h)} =  dx^\mu \, B^{(h)}_\mu (x, \theta, \bar\theta) + d\,\theta\,
\bar F^{(h)} (x, \theta, \bar\theta) + d\,\bar\theta\, F^{(h)} (x, \theta, \bar\theta) $ and the 
superfields $\Psi (x, \theta, \bar\theta)$ and $\bar\Psi (x, \theta, \bar\theta)$ have the 
following expansions along the Grassmannian directions $(\theta, \bar\theta)$ of the 
appropriately chosen (4, 2)-dimensional supermanifold (see, e.g. [17])
\begin{eqnarray}
&& \Psi (x, \theta, \bar\theta) = \psi (x) + i\,\theta\, \bar b_1 (x) + i\,\bar\theta\, b_1 (x) 
+ i\,\theta\bar\theta\, f(x), \nonumber\\
&& \bar\Psi (x, \theta, \bar\theta) = \bar\psi (x) + i\,\theta\, \bar b_2 (x) + i\,\bar\theta\, 
b_2 (x) + i\,\theta\bar\theta\, \bar f(x). 
\end{eqnarray}
Here $b_1 \equiv b_1 \cdot T,\, b_2 \equiv b_2 \cdot T,\,\bar b_1 \equiv \bar b_1 \cdot T,\,
\bar b_2 \equiv \bar b_2 \cdot T$ are the bosonic secondary fields in the above expansions
and $f \equiv f \cdot T,\, \bar f \equiv \bar f \cdot T$ are the fermionic secondary fields.
It is obvious that the above superfields are the generalizations of the ordinary 4$D$  fermionic Dirac 
fields $\psi (x)$ and $\bar \psi (x)$ onto the (4, 2)-dimensional supermanifold because, in the limit 
$\theta = \bar\theta = 0$, we retrieve the latter fields from the above superfield expansions.
We also note that the r.h.s. of the GIR (7) is a gauge-invariant quantity. Furthermore, 
the decisive feature of GIR (7) is the key observation that this relationship blends together 
the ideas of HC and GIR in a meaningful manner where the expansions of $\bigl( B^{(h)}_\mu 
(x, \theta, \bar\theta),\, F^{(h)} (x, \theta, \bar\theta), \,\bar F^{(h)} (x, \theta, \bar\theta)
\bigr)$ (cf. (6)) play very important role (as the super 1-form $\tilde{A}^{(1)}_{(h)} \, = \, dx^\mu\, 
B^{(h)}_\mu \,(x, \theta, \bar\theta)\, +\, d\,\theta\,\bar F^{(h)} \,(x, \theta, \bar\theta) \,+\, d\,\bar\theta\, 
F^{(h)} (x, \theta, \bar\theta)$ contains these results from the HC and it appears on the l.h.s. of the 
GIR (7)). Thus, we note that (7) incorporates the results of HC.

Taking into account the expansions from (8) and the expressions obtained after the
application of the HC (cf. (6)), we obtain the following relationships [17] between 
the secondary fields of (8) and the basic and auxiliary fields of the Lagrangian 
densities ${\cal L}_B$ and ${\cal L}_{\bar B}$, namely;
\begin{eqnarray}
 b_1 &=& -\, (C \cdot T)\, \psi, \qquad\qquad\quad \bar b_1 
\;= \; -\, (\bar C \cdot T)\, \psi, \nonumber\\
b_2 &=& -\,\bar\psi\, (C \cdot T), \qquad\qquad\quad \bar b_2 
\;= \;-\,\bar\psi\, (\bar C \cdot T), \nonumber\\
f &=& -\,i\,  \bigl[B + \bar C \, C\bigr]\,\psi, 
\qquad\quad \bar f \;= \;i\, \bar\psi\, \bigl[B + C \,\bar C \bigr].
\end{eqnarray}
It will be noted that the above values of $f$ and $\bar f$ are  slightly different from 
such results obtained in [15] (for the values 
of $f$ and $\bar f$). However, our present results and relationships in (9) are correct. 
The substitution of the above values of the secondary fields into the expansions (8)
yields the following
\begin{eqnarray}
\Psi^{(g)} (x, \theta, \bar\theta) &=& \psi (x) + \theta\, (-\,i\,\bar C \cdot T) \,\psi
+ \bar\theta\, (-\,i\,C \cdot T)\, \psi + \theta\bar\theta\, \left( B +  \bar C\,C
\right)\,\psi \nonumber\\
&=& \psi (x) + \theta\, (s_{ab}\,\psi) + \bar\theta\,(s_b\, \psi) + \theta\bar\theta\,
(s_b\,s_{ab}\,\psi), \nonumber\\
\bar\Psi^{(g)} (x, \theta, \bar\theta) &=& \bar\psi (x) + \theta\left[\bar\psi\,(-\,i\,\bar C \cdot T)\right] 
+ \bar\theta\, \left[\bar\psi\,(-\,i\, C \cdot T)\right] + \theta\bar\theta\,\left[\bar\psi\, 
\left(-\,B - C \,\bar C \right)\right] \nonumber\\
&=& \bar\psi (x) + \theta\, (s_{ab}\,\bar\psi) + \bar\theta\,(s_b\, \bar\psi)  + \theta\bar\theta\,
(s_b\,s_{ab}\,\bar\psi),
\end{eqnarray}  
where the superscript $(g)$ denotes the superfield expansions after the application of 
GIR. Thus, we observe that the coefficients of $\theta, \bar\theta$ and $\theta\bar\theta$  
yield the symmetry transformations corresponding to $s_{ab}, s_b$ and $s_b\,s_{ab}$, 
respectively. In other words, we have obtained a relationship between the translational
generators $(\partial_\theta, \,\partial_{\bar\theta})$ along the Grassmannian directions
$(\theta, \bar\theta)$ of the (4, 2)-dimensional supermanifold {\it and} the (anti-)BRST 
transformations $s_{(a)b}$ for the matter fields $\psi$ and $\bar\psi$ in the 
ordinary 4$D$ spacetime (cf. (2)). The appropriate mapping between these quantities of interest is: 
$ s_b\,\psi = \partial_{\bar\theta}\,\Psi^{(g)}(x, \theta, \bar\theta)|_{\theta = 0},  
\,\; s_{ab}\, \psi = \partial_\theta\,\Psi^{(g)} (x, \theta, \bar\theta)|_{\bar\theta = 0}$
and $s_b\,s_{ab}\,\psi = \partial_{\bar\theta}\,\partial_\theta\, \Psi^{(g)} (x, \theta, 
\bar\theta)$.

\noindent
\section{SUSP Unitary Operator: Key Consequences}

It is clear from the expansions (10) that we have already derived the (anti-) BRST symmetry 
transformations $s_{(a)b}$ for the Dirac fields which turn out to be the coefficients of
$ (\theta)\bar\theta $ in these expansions. Furthermore, the coefficients of $(\theta\,\bar\theta)$
turn out to be $(s_b\,s_{ab}\,\psi)$ and  $(s_b\,s_{ab}\,\bar\psi)$, respectively. These
expansions  (cf. (10)) can be re-expressed as follows 
\begin{eqnarray}
\Psi^{(g)} (x, \theta, \bar\theta) &=& \Bigl[ 1  + \theta\,(-i\,\bar C ) + \bar\theta\,(-\,i\,C )
 + \theta\bar\theta \left(B +  \bar C \, C\right)\Bigr]\,\psi (x) \nonumber\\
&\equiv&  U (x, \theta,\bar\theta) \,\psi (x), \nonumber\\
\bar\Psi^{(g)} (x, \theta,\bar\theta) &=& \bar\psi (x) \,\Bigl[ 1  + \theta\,(i\,\bar C ) 
+ \bar\theta\,(i\,C ) + \theta\bar\theta \left(-B  - C \, \bar C \right)\Bigr] \nonumber\\
&\equiv&  \bar\psi (x) \,U^\dagger (x, \theta,\bar\theta),
\end{eqnarray}
where all the fields are defined in the Lie-algebraic space (e.g. $C = C \cdot T,\,
\bar C = \bar C \cdot T,\, B = B \cdot T,\,\bar B = \bar B \cdot T$). Thus, it is obvious 
that we have derived the SUSP unitary operator $U (x, \theta,\bar\theta) $ and its Hermitian 
conjugate $U^\dagger (x, \theta,\bar\theta)$ in a very natural fashion. It is 
elementary to check that the criterion for 
the unitarity condition is satisfied by the above SUSP operators:
\begin{equation}
U (x, \theta,\bar\theta)\,U^\dagger (x, \theta,\bar\theta)\, = \,U^\dagger (x, \theta,\bar\theta)
\,U (x, \theta,\bar\theta) = 1.
\end{equation}
We emphasize that we have taken into account the fermionic nature (i.e. $\theta\,C + C\,\theta = 0,\,
\bar C\,\bar\theta + \bar\theta\,\bar C = 0,\, \theta\,\bar C + \bar C\,\theta = 0,$ etc.) of the 
Grassmannian variables $(\theta, \bar\theta)$ and (anti-)ghost fields $(\bar C)C$ in the above proof of unitarity.  
The above SUSP unitary operators can be exponentiated, in a mathematically precise fashion, 
as follows:
\begin{eqnarray}
&& U (x, \theta,\bar\theta) = exp \Bigl[  \theta\,(-i\,\bar C \cdot T) 
+ \bar\theta\,(-\,i\,C \cdot T) + \theta\bar\theta \left(B \cdot T +
 \frac{(C \times \bar C)}{2} \cdot T\right)\Bigr], \nonumber\\
&& U^\dagger (x, \theta,\bar\theta) = exp \Bigl[ \theta\,(i\,\bar C \cdot T) 
+ \bar\theta\,(i\,C \cdot T)  + \theta\bar\theta \left(-B \cdot T  - \frac{(C \times \bar C)}{2} 
\cdot T\right)\Bigr].
\end{eqnarray}
We mention that $\{C, \, \bar C\} = (C \times \bar C)$ has been taken into account 
due to $[T^a, \, T^b] = f^{abc} \, T^c$ and the
fermionic ($C^a \, {\bar C}^b + {\bar C}^b \, C^a = 0 $) nature of the (anti-)ghost fields $(\bar C)C$.
The above exponential form of the unitary operator respects the $SU (N)$ group structure in the transformation 
superspace because:
\begin{eqnarray}
\psi (x) \,&\to &  \,\Psi^{(g)} (x, \theta, \bar\theta) 
= exp \Bigl[\theta\,(-i\,\bar C ) + \bar\theta\,(-i\,\,C) +
  \theta\bar\theta \left(B +  \frac{(C \times \bar C)}{2} \right)\Bigr]\,\psi (x) \nonumber\\
&\equiv& U (x, \theta, \bar\theta)\, \psi (x), \nonumber\\
\bar\psi (x) \,&\to & \,\bar\Psi^{(g)} (x, \theta, \bar\theta) 
= \bar\psi (x) \; exp \Bigl[  \theta\,(i\,\bar C ) 
+ \bar\theta\,(i\,C)  
+ \theta\bar\theta \left(-B  - \frac{(C \times \bar C)}{2} \right)\Bigr] \nonumber\\
&\equiv& \bar\psi (x) \,  U^\dagger (x, \theta, \bar\theta).
\end{eqnarray}
In other words, we note that the SUSP unitary operators $U (x, \theta,\bar\theta)$ and
$ U^\dagger (x, \theta,\bar\theta)$ generate the shift transformations (cf. Eq. (10)) along the 
Grassmannian directions $\theta$ and $\bar\theta$ because we note that the ordinary 
fields $\psi (x)$ and $\bar\psi (x)$ are the limiting cases of the superfields 
$\Psi^{(g)} (x, \theta, \bar\theta) $ and $\bar\Psi^{(g)} (x, \theta, \bar\theta) $ 
when the limit is taken to be $\theta = \bar\theta  = 0 $ in the equation (14).

The relationship $\psi (x) \,\longrightarrow \, \Psi^{(g)} (x, \theta,\bar\theta) 
= U (x, \theta,\bar\theta) \psi (x)$, where the $SU(N)$ group structure is maintained
in the transformation superspace, allows us to define the covariant derivative in terms of the 
super exterior derivative and super 1-form connection with the following inherent property
for the 4$D$ non-Abelian 1-form $SU(N)$ gauge theory, namely;
\begin{equation}
D\,\psi (x) \,\longrightarrow\, \tilde D\, \Psi^{(g)} (x, \theta,\bar\theta) 
= U (x, \theta,\bar\theta)\, D\psi (x),
\end{equation}
where $D\psi (x) = (d + i\, A^{(1)} (x) )\,\psi (x) $ and ${\tilde D} \Psi^{(g)} (x, \theta,\bar\theta) 
= \bigl({\tilde d} + i\, {\tilde A}^{(1)}_{(h)} (x, \theta,\bar\theta) \bigr)\,
\Psi^{(g)} (x, \theta,\bar\theta) $. 
The above equation (15) leads to the following explicit transformation rule for the connection 
super 1-form in the superspace:
\begin{eqnarray}
{\tilde A}^{(1)}_{(h)} \,(x, \theta,\bar\theta) = U (x, \theta,\bar\theta)\,A^{(1)} (x) \,
{U^\dagger} (x, \theta,\bar\theta) + i\, \bigl({\tilde d}\,U (x, \theta,\bar\theta)\bigr)\,
{U^\dagger} (x, \theta,\bar\theta).
\end{eqnarray}
The substitution of the expressions for $U (x, \theta,\bar\theta)$ and ${U^\dagger} 
(x, \theta,\bar\theta)$, from (11), leads to the following explicit results from the {\it second}
term of the  r.h.s. of (16), namely:
\begin{eqnarray}
&& dx^\mu \,\left[ \theta\,(\partial_\mu \, \bar C) + \bar\theta\,(\partial_\mu C)
+ i\,\theta\bar\theta\,\left(\partial_\mu\,B + \partial_\mu\,C \times \bar C \right) 
\right] \nonumber\\
+ && d\theta\, \left[\bar C + \theta \,\left( -\,\frac{i}{2}\,(\bar C \times \bar C)\right) 
+ \bar\theta\,(i\,B) + \theta\bar\theta\,\left([B,\,\bar C]\right)\right] \nonumber\\
+ && d\bar\theta\,\Bigl[ C + \theta\,\bigl(-i\,B - \,i\, C \times \bar C\bigr) + \bar\theta\,
\left( -\,\frac{i}{2}\,(C \times C)\right) \nonumber\\  + && \theta\bar\theta\,\left(\bigl[ B 
+ C \times \bar C,\,C\bigr]\right) \Bigr].
\end{eqnarray}
Comparing with the coefficients of $(d\theta)$ and $(d\bar\theta)$ [that are present on the 
l.h.s. of the super 1-form connection $[ {\tilde A}^{(1)}_{(h)}\, =\,  dx^\mu\,\, B^{(h)}_\mu 
(x, \theta,\bar\theta) \, + \, d\,\theta\,\, {\bar F}^{(h)} (x, \theta,\bar\theta) \, + \, d\,\bar\theta \,\,
F^{(h)} (x, \theta,\bar\theta)$], we observe that we have already derived the 
{\it correct} expressions for the superfields $F^{(h)} (x, \theta,\bar\theta)$ and 
${\bar F}^{(h)} (x, \theta,\bar\theta)$ where, as is evident, $F^{(h)} (x, \theta,\bar\theta)$ 
and ${\bar F}^{(h)}(x, \theta,\bar\theta)$ are the expansions that have been obtained 
after the application of the HC. A close look at the coefficients of $d\bar\theta$ in 
(17) shows that, if we use the CF-condition: $ B + \bar B + (C \times \bar C) = 0 $, 
we have an alternative form of the last entry in (17), namely;
\begin{eqnarray}
d\bar\theta \left[ C + \theta\,(i\bar B) + \bar\theta\,\left(- \frac{i}{2}\,(C \times C)
\right) + \theta\bar\theta \bigl(-\,\bar B \times C \bigr)\right]. 
\end{eqnarray}
This expression (when compared with $d\bar\theta\, F^{(h)} (x, \theta,\bar\theta)$
from the l.h.s.) yields the transformations $s_b\,C (x),\, s_{ab} \,C(x)$ and $s_b\,s_{ab}\,C (x)$
as the coefficients of $\theta, \bar\theta$ and $\theta\bar\theta$, respectively. 
To fully derive the results of HC, we have to explicitly compute the coefficients of 
$(dx^\mu)$ from the r.h.s. of (16). The coefficient of $dx^\mu$, from the sum of the 
first term and second term on the r.h.s. of transformations (16), yields the following:
\begin{eqnarray}
dx^\mu\, \bigl[A_\mu (x) + \theta\, (D_\mu\, \bar C) + \bar\theta\, (D_\mu \,C) 
 + i\,\theta\bar\theta\, (D_\mu\, B + D_\mu C \times \bar C) \bigr],
\end{eqnarray}
where the above equation is nothing but the sum of the following:
\begin{eqnarray}
&&dx^\mu \bigl(A_\mu + i\,\theta [A_\mu, \,\bar C] + i\,\bar\theta [A_\mu,\,C] 
+ \theta\bar\theta\,\bigl(-\,[A_\mu,\,B] - \lbrace[A_\mu,\,C], \bar C\rbrace\bigr)\bigr),
\end{eqnarray}
\begin{eqnarray}
dx^\mu \, \left[\theta\; (\partial_\mu \bar C) + \bar\theta\;(\partial_\mu C)
+ i\,\theta\bar\theta\;\bigl(\partial_\mu B + \partial_\mu C \times \bar C\bigr)\right].
\end{eqnarray}
We note that the contributions (20) and (21) come out explicitly from the first and second
terms on the r.h.s. of equation (16). When the above expression is compared with the l.h.s. 
of the definition 
${\tilde A}^{(1)}_{(h)} = dx^\mu\, B^{(h)}_\mu (x, \theta,\bar\theta)
+ d\theta\, {\bar F}^{(h)} (x, \theta,\bar\theta) + d\bar\theta \,
F^{(h)} (x, \theta,\bar\theta)$, we obtain the expansion for $ B^{(h)}_\mu 
(x, \theta,\bar\theta)$ that has been derived due to HC in our earlier equation (6).

The transformations on $\tilde{F}^{(2)} = \bigl[(dZ^M \,\wedge \,dZ^N)/{2 !} \bigr] 
\tilde{F}_{MN} \,(x, \theta,\bar\theta)$ in the superspace can  also be computed by 
establishing a connection between $\tilde{F}^{(2)}$ and ${F}^{(2)} = \bigl[(dx^\mu \,\wedge 
\, dx^\nu)/{2 !} \bigr]\, {F}_{\mu\nu} (x)$. In this endeavour, the property of the 
successive operations of a couple of covariant derivatives and their connection with the curvature 
2-form plays an important role. For instance, in the ordinary 4$D$ Minkowskian flat spacetime, 
we know that:
\begin{eqnarray}
DD\,\psi (x) =  i\,F^{(2)} (x)\,\psi (x), \qquad D = d + i\, A^{(1)}.
\end{eqnarray}
This relation can be generalized onto (4, 2)-dimensional supermanifold as
\begin{eqnarray}
D\,D\,\psi \; &\longrightarrow &\; \tilde{D}\, \tilde{D}\,\Psi^{(g)} (x, \theta,\bar\theta) 
= i\, \tilde{F}^{(2)} (x, \theta,\bar\theta)\,\Psi^{(g)} (x, \theta,\bar\theta),
\end{eqnarray}
where $\tilde{D} = \tilde{d} + i\,A^{(1)}_{(h)} (x, \theta,\bar\theta)$ and 
$\Psi^{(g)} (x, \theta,\bar\theta) = U (x, \theta,\bar\theta) \psi (x)$. Actually,
the relationships of the kind in (22) and (23) are covariant relations 
[due to $\Psi^{(g)} (x, \theta,\bar\theta) =  U (x, \theta,\bar\theta)\, \psi (x) $ and
$\tilde{D}\,\Psi^{(g)} (x, \theta,\bar\theta) = U (x, \theta,\bar\theta)\, D\psi (x)$].
Thus, we have the following explicit relationship:
\begin{eqnarray}
\tilde{D}\tilde{D}\,\Psi^{(g)} (x, \theta,\bar\theta) = i\,\tilde{F}^{(2)}(x)\,U (x, \theta,\bar\theta)\,\psi (x)  
\equiv \,  i\, U (x, \theta,\bar\theta)\, F^{(2)}(x)\,\psi (x).
\end{eqnarray} 
The above equation emerges out from the equation (23) [with $\tilde{D} = U(x, \theta,\bar\theta)\\D\,
U^\dagger (x, \theta,\bar\theta)$ and $\Psi^{(g)} (x, \theta,\bar\theta) 
= U (x, \theta,\bar\theta)\,\psi (x)$]. From the relation (24), it is clear that we 
have the following explicit relationship
\begin{eqnarray}
\tilde{F}^{(2)} (x, \theta,\bar\theta) = U (x, \theta,\bar\theta)\,F^{(2)}(x)\,
U^\dagger (x, \theta,\bar\theta),
\end{eqnarray}
which has been mentioned earlier, too, as the transformation property of the 
curvature $F^{(2)}(x)$ in the superspace.

There is an altogether different theoretical method to derive the transformation
property of $F^{(2)} (x)$ in the superspace by exploiting the celebrated Maurer-Cartan 
type of equation. It will be noted that the {\it exact} Maurer-Cartan equation is 
somewhat different from the relationship that defines the curvature 2-form. 
However, it looks similar in appearance. From our earlier equation (16), 
it is evident that we have the following
\begin{eqnarray}
\tilde{d}\,\tilde{A}^{(1)}_{(h)} (x, \theta,\bar\theta) = (\tilde{d}U)\,A^{(1)}(x)\,U^\dagger 
+ U\,dA^{(1)} (x)\,U^\dagger - U\,A^{(1)}(x)\,\tilde{d}U^\dagger - i\, (\tilde{d}U)(\tilde{d}U^\dagger), 
\end{eqnarray}
where we have used the properties $\tilde{d}(UU^\dagger) = 0\; \Longrightarrow\;
(\tilde{d}U)\,U^\dagger = - \,U\,(\tilde{d}U^\dagger),\,\tilde{d}^2 = 0$ and 
$\tilde{d}\,A^{(1)} (x) =  dA^{(1)} (x)$.
We further observe, from (16), that
\begin{eqnarray}
i\,\tilde{A}^{(1)}_{(h)}(x, \theta,\bar\theta) \,\wedge\,\tilde{A}^{(1)}_{(h)} (x, \theta,\bar\theta)
&=& U\,\bigl(i\,A^{(1)} (x)\,\wedge \,A^{(1)}(x)\bigr)\,U^\dagger + U\,A^{(1)}(x)\,(\tilde{d}U^\dagger)\nonumber\\  
&-& (\tilde{d}U)\,A^{(1)}(x)\,U^\dagger + i\,(\tilde{d}U)(\tilde{d}U^\dagger),
\end{eqnarray}
where, once again, we have used the inputs from the differential geometry (i.e. $\tilde{d}^2 = 0$) 
and $(\tilde{d}\,U)U^\dagger + U\,(\tilde{d}U^\dagger) = 0$. The sum of (26) and (27) yields:
\begin{eqnarray}
\bigl(\tilde{d}\,\tilde{A}^{(1)}_{(h)} + i\, \tilde{A}^{(1)}_{(h)}\,\wedge\,\tilde{A}^{(1)}_{(h)}\bigr) = 
U (x, \theta,\bar\theta)\,\Bigl[dA^{(1)}(x) + i\,A^{(1)}(x)\,\wedge\,A^{(1)}(x)\Bigr]\,U^\dagger 
(x, \theta,\bar\theta).
\end{eqnarray}
The above relation is basically the relationship
$\tilde{F}^{(2)} (x, \theta,\bar\theta) = U (x, \theta,\bar\theta)\,
F^{(2)}(x) \,U^\dagger (x, \theta,\bar\theta)$ that has been derived in (25) by exploiting the 
property of covariant derivatives. Ultimately, we conclude that there are, at least, two different 
and distinct theoretical methods to 
compute the relationships ($\tilde{F}^{(2)} (x, \theta,\bar\theta) = U (x, \theta,\bar\theta)
\,F^{(2)}(x)\,U^\dagger \,(x, \theta,\bar\theta)$) between $\tilde{F}^{(2)} (x, \theta,\bar\theta)$ 
and $F^{(2)} (x)$ which are the super and ordinary curvature 2-forms in our (4, 2)-dimensional SUSY and 
(3 +1)-dimensional ordinary 
theories, respectively.

We would like to briefly comment on the relationships between the GIR (invoked in equation (7)) and
the SUSP operators (that have been derived in our earlier equations (11) and (13)). Towards this goal in mind, we 
focus on the basic definitions of the (super)covariant derivatives that have been quoted in (15). 
In fact, in its explicit form, this equation can be written as:
\begin{eqnarray}
(\tilde d + i\,\tilde A^{(1)}_{(h)} ) \, \Psi (x, \theta,\bar\theta) 
=  U (x, \theta,\bar\theta) \, (d + i\, A^{(1)} (x)) \, \psi (x).
\end{eqnarray}
Using the fundamental definitions of $\tilde d$ and $\tilde A^{(1)}_{(h)}$ from the equations 
(3) and (16), we obtain the following equalities when we compare the l.h.s. with the r.h.s. of
the above equation
\begin{eqnarray}
&& dx^\mu \Bigl[ \partial_\mu \, \Psi + i\, U\, A_\mu (x) \, U^\dagger  \Psi - 
(\partial_\mu U) U^\dagger  \Psi\Bigr] = dx^\mu  \bigl(U \partial_\mu  \psi (x) 
+ i\, U\, A_\mu  \psi(x)\bigr), \nonumber\\
&& d\,\theta \,\Bigl[\partial_\theta \, \Psi(x, \theta,\bar\theta) - 
(\partial_\theta \, U)\, U^\dagger \, \Psi(x, \theta,\bar\theta)\Bigr] = 0,\nonumber\\
&& d\,\bar\theta \,\Bigl[\partial_{\bar\theta} \, \Psi(x, \theta,\bar\theta) - 
(\partial_{\bar\theta} \, U)\, U^\dagger \, \Psi(x, \theta,\bar\theta)\Bigr] = 0,
\end{eqnarray}
where we have used the abbreviations $U \equiv  U(x, \theta,\bar\theta) $ and $U^\dagger \equiv 
 U^\dagger (x, \theta,\bar\theta) $.
It is quite elementary to observe that the solution: $\Psi^{(g)} (x, \theta,\bar\theta) =
U(x, \theta,\bar\theta) \, \psi (x)$ does satisfy the top two equations in (30). It is 
interesting to note that this solution {\it also} satisfies the last equation that is present in (30). 
If we substitute the exact form of $U(x, \theta,\bar\theta) $ from (11) into this relationship, 
we obtain the result (10) which are derived due to the application of GIR. We further point out that
the solution $\Psi^{(g)} (x, \theta,\bar\theta) = U(x, \theta,\bar\theta) \, \psi (x)$ implies that
we {\it also} have $\bar\Psi^{(g)} (x, \theta,\bar\theta) = \bar\psi (x) \,U^\dagger(x, \theta,\bar\theta)$.
If we substitute the exact mathematical expression for $U^\dagger(x, \theta,\bar\theta)$ from 
(11) into this relationship, we obtain the (anti-)BRST symmetry transformations for $\bar \psi (x)$
field as quoted in (10). Ultimately, we lay emphasis on our key observation that the exact mathematical
derivation of the SUSP operators $U$ and  $U^\dagger$ provide, in some sense, the alternative to HC 
as well as GIR (where the $SU(N)$ group structure of the non-Abelian theory is very explicitly maintained
in the superspace).

Before we end this section, we would like to lay emphasis on our observation that we have derived the
SUSP operators $U  (x, \theta,\bar\theta)$ and  $U^\dagger  (x, \theta,\bar\theta)$ without invoking any
Hermitian conjugation condition from {\it outside} on the parameters ($\theta, \,\bar\theta$) and fields
($C, \, \bar C, \, B$) of our SUSY theory. However, it can be checked explicitly that if we apply the following
Hermitian conjugation operations from outside, namely;
\begin{eqnarray}
&&\theta^\dagger = \mp \;\theta, \qquad \bar\theta^\dagger = \mp \;\bar\theta, 
\qquad C^\dagger = \pm \;C,\quad \bar C^\dagger 
= \pm \;\bar C, \qquad B^\dagger = B, \nonumber\\ 
&&   i^\dagger = - i, \qquad (\theta\,\bar\theta)^\dagger = {\bar\theta}^\dagger \, \theta^\dagger 
= -\, \theta\,\bar\theta, \qquad (C\,\bar C)^\dagger = {\bar C}^\dagger \, C^\dagger = \,\bar C\, C,  
\end{eqnarray} 
the SUSP unitary operators interchange (i.e. $U  (x, \theta,\bar\theta)
\longleftrightarrow U^\dagger  (x, \theta,\bar\theta)$). In other words, by using the 
above operations, we can obtain $U^\dagger  (x, \theta,\bar\theta)$ from $U (x, \theta,\bar\theta)$
and {\it vice-versa}. Thus, our theory does support a set of Hermitian conjugate symmetry
 transformations, too. It should be noted that the Hermitian conjugation conditions,
quoted in (31), are {\it not} unique. Our theory might support other set of Hermitian 
conjugation conditions, as well.

\noindent
\section{Conclusions}

In our present investigation, we have explicitly computed the mathematically precise form of the 
SUSP unitary operator $U (x, \theta,\bar\theta)$ that appears in the transformation 
superspace of the Dirac field $\psi (x)\,\rightarrow\,\Psi^{(g)} (x, \theta,\bar\theta)
= U (x, \theta,\bar\theta) \psi (x)$. We have {\it also} 
computed the Hermitian conjugate SUSP operator (i.e. $U^\dagger (x, \theta,\bar\theta)$)
that appears in the transformation of the Dirac field $\bar\psi (x) \, 
\rightarrow \,\bar\Psi^{(g)} (x, \theta,\bar\theta) = \bar\psi (x)\,U^\dagger 
(x, \theta,\bar\theta)$. We have shown that $U\,U^\dagger =  U^\dagger\, U = 1$ which
proves the unitarity of this SUSP operator. The key point, to be noted, is the observation
that $U (x, \theta,\bar\theta)$ and  $U^\dagger (x, \theta,\bar\theta)$ appear very naturally
from the transformation properties of $\psi (x)$ and $\bar\psi (x)$ to 
$\Psi^{(g)} (x, \theta,\bar\theta)$ and  $\bar\Psi^{(g)} (x, \theta,\bar\theta)$, respectively. 
We have {\it not} invoked any Hermitian conjugation operation  {\it from outside} on the parameters
($\theta, \bar\theta$)
and fields ($ C, \bar C, B$, etc.) of our theory for the computation of 
$U^\dagger (x, \theta,\bar\theta) $ from the known expression for $U (x, \theta,\bar\theta)$.
However, to demonstrate that our theory {\it also} supports a set of Hermitian conjugation operations,
we have listed a couple of such operations in equation (31) on the parameters 
as well as appropriate fields of our theory which provide
a connection between $U (x, \theta,\bar\theta)$ and $U^\dagger (x, \theta,\bar\theta)$.

The above SUSP transformations (which maintain the $SU(N)$ group structure) lead to 
the definition of the covariant derivative on the (4, 2)-dimensional supermanifold. 
This definition, in turn, enables us  to obtain the transformation ($ A^{(1)}\,\rightarrow\,
\tilde{A}^{(1)}_{(h)} $) on the 1-form connection (cf. (16)). This relationship immediately 
leads to the derivation of the results of HC. In other words, we derive the expansions (6) 
in terms of the SUSP operators $U (x, \theta,\bar\theta)$ and $U^\dagger (x, \theta,\bar\theta)$
due to the appearance of equation (16) in our present endeavor. Thus, the precise 
derivation of SUSP unitary operators $U$ and $U^\dagger$ provides an alternative to
the HC where the group structure is maintained {\it explicitly} in the superspace. 
In fact, these SUSP operators have also enabled us to obtain the results of GIR where, 
once again, the idea of covariant derivative has played an important role. We have 
discussed the derivation of the relationships: $\Psi^{(g)} (x, \theta,\bar\theta)
= U (x, \theta,\bar\theta)\, \psi (x), \,\,\, \bar\Psi^{(g)} (x, \theta,\bar\theta)
\,= \,\bar\psi (x)\,\,U^\dagger (x, \theta,\bar\theta)$ in our equations (29) and (30) (cf. Sec. 5)
where the mathematical  forms of SUSP operators $U (x, \theta,\bar\theta)$ and 
$U^\dagger (x, \theta,\bar\theta)$ play crucial roles.

As we know, two successive applications of the covariant derivative 
$D = (d + i\, A^{(1)})$ on the Dirac field $\psi (x)$ (i.e. $DD\,\psi$) leads to the 
definition of the curvature 2-form $ F^{(2)}$ (i.e. $DD\,\psi = i\,F^{(2)}\,\psi $). 
This relation is a covariant relation and one can tap the potential of this relationship
to obtain its SUSY version (i.e. $DD\,\psi \;\longrightarrow\; \tilde D\,\tilde D\,
\Psi^{(g)} = i\,{\tilde F}^{(2)} \,\Psi^{(g)}  $). Now, using the well-known relationship
of the covariant derivative $\tilde D\, \Psi^{(g)} = U\,D\,\psi$, we obtain the relationship  
${\tilde F}^{(2)}\, U (x, \theta,\bar\theta) = U (x, \theta,\bar\theta) F^{(2)} (x)
\Longrightarrow  {\tilde F}^{(2)} (x, \theta,\bar\theta) = U (x, \theta,\bar\theta)\,
F^{(2)}(x)\,U^\dagger (x, \theta,\bar\theta)$ which leads to the derivation of the 
(anti-)BRST transformations for the curvature tensor $F_{\mu\nu}$ (i.e. $s_b \, F_{\mu\nu} 
= i\, (F_{\mu\nu} \times C)$ and $s_{ab} \, F_{\mu\nu} = i\, (F_{\mu\nu} \times \bar C)$)
in the context of our 4D {\it non-Abelian} theory. The above precise relationship 
(i.e. $\tilde{F}^{(2)} (x, \theta,\bar\theta) = U (x, \theta,\bar\theta)\,
F^{(2)}(x)\,U^\dagger (x, \theta,\bar\theta) $) reduces to its {\it Abelian} counterpart 
$\tilde{F}^{(2)} (x, \theta,\bar\theta) = F^{(2)} (x)$ that has been discussed in our earlier 
work [11] on the interacting Abelian $U (1)$ gauge theory with Dirac and complex scalar fields
where there is an explicit coupling between the gauge field and matter fields.

In our earlier works [18-20], we have derived the nilpotent (anti-)BRST and (anti-)co-BRST symmetry 
transformations for the Stueckelberg-modified version of the 2D Proca theory [18], the modified 
version of the 2D anomalous gauge theory [19] and the 2D self-dual chiral bosonic field theory [20] 
by exploiting the tools and techniques of the augmented version of geometrical superfield 
formalism [8-10]. We lay emphasis on the fact that, in these theories [18-20], we have the presence 
of the matter as well as gauge fields which are coupled to one-another in a specific fashion. 
Thus, it would be very nice idea to find out the SUSP unitary operators for these theories where, 
not only the off-shell nilpotent (anti-)BRST symmetry transformations 
but the off-shell nilpotent (anti-)co-BRST symmetries exist, too, for the matter,  gauge 
and (anti-)ghost fields. We are currently intensively involved with these ideas. In this connection, it is gratifying
to state that we have partially accomplished this goal in our very recent work [21] where we have discussed {\it only}
about the (anti-)BRST symmetries and corresponding SUSP unitary operators $U (x, \theta,\bar\theta)$ 
and $U^\dagger (x, \theta,\bar\theta)$. The discussion about the nilpotent (anti-)co-BRST symmetry transformations
and the corresponding {\it dual} SUSP unitary operators has been achieved in our very recent work [22].
In this context, it is very gratifying to state that we have also derived the {\it dual}
SUSP unitary operator and its Hermitian conjugate which lead to the derivation
of the proper (anti-)co-BRST symmetry transformations for some of the interesting
Abelian models [18-20] of the Hodge theory.\\

\vskip .5 cm

\noindent
{\bf Acknowledgements}\vskip .65 cm
\noindent
One of us (RPM) would like to express his deep sense of gratitude to the AS-ICTP
and SISSA, Trieste, Italy for the warm hospitality extended to him during his participation in the 
conference on {\bf ``Aspects of Gauge and String Theories" (1 - 2 July 2015)} which was held at SISSA to
celebrate the {\bf $70^{th}$   birth anniversary of L. Bonora}. The idea behind this work came
during this conference. D. Shukla is thankful to the UGC, Government 
of India, New Delhi, for financial support through RFSMS-SRF scheme and T. Bhanja is grateful to 
the BHU-fellowship under which the present investigation has been carried out. Fruitful
comments by our esteemed {\bf Reviewer} are gratefully acknowledged, too.  \\

\vskip .8 cm 

\begin{center}
\Large{\bf Appendix A: More on the (anti-)BRST invariant Lagrangian densities
in the Curci-Ferrari gauge}\\
\end{center}
\vskip .8 cm
For the readers' convenience,
we discuss explicitly a few more key points, connected with 
the form of our Lagrangian densities ${\cal L}_B$ and ${\cal L}_{\bar B}$ in the Curci-Ferrari
gauge [cf. Eq. (1)] of our present investigation 
(where we have already taken $\xi = 2$). 
It is very interesting to note that (modulo some total
spacetime derivatives)
\begin{eqnarray}
&&{\cal L}_B = - \frac{1}{4}\, F_{\mu\nu} \cdot F^{\mu\nu} + \bar \psi\, (i\,\gamma^\mu\,D_\mu - m)\, 
\psi + s_b\,s_{ab} \Bigl(\frac{i}{2}\, A_\mu \cdot A^\mu - \bar C \cdot C\Bigr),\nonumber\\
&&{\cal L}_{\bar B} = - \frac{1}{4}\, F_{\mu\nu} \cdot F^{\mu\nu} + \bar \psi\, (i\,\gamma^\mu\,D_\mu - m)\, 
\psi - s_{ab}\,s_{b} \Bigl(\frac{i}{2}\, A_\mu \cdot A^\mu - \bar C \cdot C\Bigr),
\end{eqnarray}
where the covariant derivative on the matter field 
[i.e. $D_\mu \,\psi = (\partial_\mu + i A_\mu \cdot T)\,\psi$] is in the fundamental 
representation of the $SU(N)$ Lie algebra. We note that the following are true:
\begin{eqnarray}
s_b\,s_{ab} \Bigl(\frac{i}{2}\, A_\mu \cdot A^\mu \Bigr) &=&
B\cdot (\partial_\mu A^\mu) -\,i\,\partial_\mu\bar C \cdot D^\mu C - 
\partial_\mu \,(A^\mu \cdot B),\nonumber\\
 - \, s_{ab}\,s_{b} \Bigl(\frac{i}{2}\, A_\mu \cdot A^\mu\Bigr) &=&
-\, \bar B \cdot (\partial_\mu A^\mu) - i D_\mu\bar C \cdot \partial^\mu C 
- \partial_\mu \,(A^\mu \cdot \bar B).
\end{eqnarray}
Here the covariant derivative $D_\mu\, C = \partial_\mu \,C + i (A_\mu \times C)$ 
is in the adjoint representation of the $SU(N)$ Lie algebra. Thus,
we note that the gauge-fixing and Faddeev-Popov ghost terms of the Lagrangian
densities ${\cal L}_B$ and ${\cal L}_{\bar B}$  are derived from (33)
modulo a total spacetime derivative term. In the above derivations, we have {\it not} used
the CF-conditions {\it anywhere}.

Now, we concentrate on the computations of the following equivalent expressions 
\begin{eqnarray}
s_b\,s_{ab} (-\, \bar C \cdot C) =  s_{ab}\,s_{b} (\bar C \cdot C),
\end{eqnarray}
where we have to use the CF-condition because the absolute anticommutativity
$(s_b\,s_{ab} + s_{ab}\,s_b) = 0$ is satisfied {\it only} on the hypersurface
where $B + \bar B + C \times \bar C = 0$ is true in the $4D$ Minkowskian 
spacetime manifold. In this context, we note that we have the following
\begin{eqnarray}
s_b\,s_{ab} (-\, \bar C \cdot C) &=&  -\, (B + \bar B)\cdot (C \times \bar C)  
- B\cdot \bar B + \frac{1}{4}\, (\bar C \times \bar C ) \cdot (C \times  C)\nonumber\\
&\equiv&   -\, (B + \bar B)\cdot (C \times \bar C)  - B\cdot \bar B 
- \frac{1}{2}\, (C \times \bar C) \cdot (C \times \bar C).
\end{eqnarray}
Now, using the CF-condition $(C \times \bar C) = -\, (B + \bar B)$, we obtain the following:
\begin{eqnarray}
s_b\,s_{ab} (- \, \bar C \cdot C) =  \frac{1}{2}\, (B \cdot B + \bar B \cdot \bar B). 
\end{eqnarray}
We observe that the gauge-fixing and Faddeev-Popov ghost terms of the
Lagrangian densities ${\cal L}_B$  and ${\cal L}_{\bar B}$ of our Eq. (1) are
nothing other the appropriate sum of (33) and (36).
It is straightforward to note that $s_{ab}\,s_{b} (\bar C \cdot C)$ produces  the 
same result as in (36) (when we use the CF-condition). The (anti-)BRST symmetry 
transformations $s_{(a)b}$ of the 
above Lagrangian densities yield the following total spacetime derivatives plus 
terms that are zero on the hypersurface where the CF-condition is valid, namely;
\begin{eqnarray}
s_b\, {\cal L}_B &=& \partial_\mu [B \cdot D^\mu C], \qquad \qquad
s_{ab}\, {\cal L}_{\bar B} = \partial_\mu [- \, \bar B \cdot D^\mu \bar C],\nonumber\\
s_{ab}\, {\cal L}_{B} &=& \partial_\mu \,[-\, \{\bar B + (C \times \bar C)\}\cdot \partial^\mu \bar C] + 
(B + \bar B + C \times \bar C) \cdot D_\mu \,(\partial^\mu \bar C)\nonumber\\ 
 &\equiv & \partial_\mu [  B \cdot \partial^\mu \bar C] + 
(B + \bar B + C \times \bar C) \cdot D_\mu \,(\partial^\mu \bar C), \nonumber\\
s_b\, {\cal L}_{\bar B}& =& \partial_\mu [\{B + (C \times \bar C)\} \cdot \partial^\mu C] -
(B + \bar B + C \times \bar C) \cdot D_\mu\,(\partial^\mu C)\nonumber\\
 &\equiv& \partial_\mu [- \bar B \cdot \partial^\mu C] -
(B + \bar B + C \times \bar C) \cdot D_\mu\,(\partial^\mu C). 
\end{eqnarray}
Thus, as far as the symmetry properties are concerned, the Lagrangian densities 
${\cal L}_B$ and ${\cal L}_{\bar B}$  are {\it equivalent} on the hypersurface 
(in the $4D$ Minkowskian spacetime manifold) where the
CF-condition $B + \bar B + C \times \bar C = 0$ is satisfied. In fact, the absolute
anticommutativity property $(s_b\,s_{ab} + s_{ab}\,s_b = 0)$ of the (anti-)BRST 
symmetry transformations $s_{(a)b}$ is also true {\it only} on this hypersurface.
We end this Appendix with the remark that the Lagrangian densities ${\cal L}_B$ and
${\cal L}_{\bar B}$ can also be expressed as:
\begin{eqnarray}
{\cal L}_B = &-& \,{1 \over 4}\, F^{\mu\nu}\cdot F_{\mu\nu} + \bar \psi\, (i\,\gamma^\mu\,D_\mu - m)\, \psi
+ B\cdot (\partial_\mu A^\mu) + B\cdot B + B \cdot (C \times \bar C) \nonumber\\
&+& {1 \over 2}\,(C \times \bar C)\cdot (C \times \bar C)  -\,i\,\partial_\mu\bar C \cdot D^\mu C, \nonumber\\
{\cal L}_{\bar B} = &-& \,{1 \over 4}\, F^{\mu\nu}\cdot F_{\mu\nu} + \bar \psi\, (i\,\gamma^\mu\,D_\mu - m)\, \psi
- \bar B\cdot (\partial_\mu A^\mu) + \bar B\cdot \bar B + \bar B \cdot (C \times \bar C)\nonumber\\
 &-& {1 \over 2}\,(C \times \bar C)\cdot (C \times \bar C) -  i D_\mu\bar C \cdot \partial^\mu C.
\end{eqnarray}
The equations of motion (i.e. ${{\partial{\cal L}_B}\over {\partial B}} = 0, \, 
{{\partial{\cal L}_{\bar B}}\over {\partial \bar B}} = 0 $) w.r.t. $B$ and $\bar B$ yield the following
\begin{eqnarray}
\partial_\mu A^\mu &+& 2 B + (C \times \bar C) = 0, \nonumber\\
-\,\partial_\mu A^\mu &+& 2 \bar B + (C \times \bar C) = 0. 
\end{eqnarray}
The sum of the above relationships leads to the derivation of the CF-condition.\\

\end{document}